\documentclass{article}

\usepackage{arxiv}

\usepackage[utf8]{inputenc} 
\usepackage[T1]{fontenc}    
\usepackage{hyperref}       
\usepackage{url}            
\usepackage{booktabs}       
\usepackage{amsfonts}       
\usepackage{nicefrac}       
\usepackage{microtype}      
\usepackage{amsmath}        
\usepackage{lipsum}		
\usepackage{graphicx}
\usepackage[numbers,sort&compress]{natbib}
\usepackage{doi}

\title{A generalized rate law for inhomogeneous system and turbulence-chemistry decoupling of reaction rate calculation in combustion}

\author{Xiang-Yuan Li\thanks{Corresponding author. Email: xyli@scu.edu.cn}, \ Xin-Yu Zhang, \ Chuanfeng Yue\\
	College of Chemical Engineering\\
    Engineering Research Center of Combustion and Cooling for Aerospace Power, Ministry of Education\\
	Sichuan University\\
	Chengdu, 610065 P.R.China
}

\date{}

\renewcommand{\shorttitle}{A generalized rate law and turbulence-chemistry decoupling for combustion}

\begin{document}
\maketitle

\begin{abstract}
	In this work, the rate law for inhomogeneous concentration distributions has been formulated, by applying spatial integration over the products of species concentrations. Reaction rates for typical reactions have been investigated by assuming a linear concentration distribution in the grid. A few examples of one-dimensional concentration distributions, straight line, piecewise, and sine function, for a selected second order reaction have been taken to illustrate the validations of the method developed. Difference between the reaction rates by spatial integration and by mean concentrations have been discussed. It is revealed that the chemical reaction rates for combustion simulation can be calculated by appropriate sub-grid modeling of concentration distributions, without needs of the explicit consideration of turbulent combustion interactions, and the reaction rates for the species transport equation in turbulent combustion simulations can be accurately calculated if the concentration distributions of species within the grid are correctly defined.
\end{abstract}


\section{Introduction}
The chemical reaction rate law can be generally expressed as the following standard form, \emph{i.e.}\citep{atkins2023atkins, jesudason2009internal}
\begin{equation}
\label{eq1}
    \frac{dc_{\mathrm A}}{dt}=\widehat{R}_{\mathrm A}=-kc_{\mathrm A}^\alpha c_{\mathrm B}^\beta\cdots
\end{equation}
where $c_i = \frac{n_i}{V}$ is the molar concentration of species $i$ and $k$ the rate constant. For an inhomogeneous system such as in turbulent combustion, Eq.(\ref{eq1}) is generally unable to give the correct rate $\frac{dc_{\mathrm A}}{dt}$. It should be noticed that parameters $alpha$ and $\beta$ in Eq.(\ref{eq1}) are generally fitted from experiments. Due to the complexity of the combustion chain reactions, it is usually a difficult task to determine the reaction order $(\alpha + \beta + \ldots)$. However, for a simple elementary reaction step, the law of mass action has been applied for the determination of the reaction order. Such a treatment is crucial in the constructions of complex reaction mechanisms such as those in combustion reactions.

From the point of view of chemical kinetics, the change of species mass fraction is expressed as
\begin{equation}
    \frac{dy_i}{dt} = \frac{M_i}{\rho}\frac{dc_i}{dt}
\end{equation}
Here $y_{i}$ is the mass fraction of species $i$, $M_{i}$ the molecular weight, and $\rho$ the density. For a system of $N$ species and $L$ elementary steps, a reaction mechanism is generally composed of a list of ``elementary'' steps. The mechanism generally includes only single-molecule (first-order) and bimolecular (second-order) reactions while the third and higher order reactions are rare according to the law of mass action. The total formation rate of species $i$ in complex reactions is called the chemical source term, \emph{i.e.},
\begin{equation}
\label{eq3}
    \omega_{i} = M_{i}\sum_{r = 1}^{L}{\widehat{R}}_{i,r}
\end{equation}
where $r$ denotes the reaction step. Eq.(\ref{eq1}) keeps true for homogeneous system in which the reactant concentrations are uniformly distributed in the entire reaction container, or in a grid in the numerical simulations of reaction flow\citep{plutschack2017hitchhiker}. However, there are often different degrees of concentration gradients and local fluctuations in the actual reaction systems. Reaction flow in combustion is one of the most complex phenomena, and one of the keys to its numerical modeling is the description of the chemical reactions of fuels\citep{atkins2023atkins, plutschack2017hitchhiker, egolfopoulos2014advances, westbrook1977numerical} The numerical simulation of turbulent combustion is based on two key components: reaction kinetics\citep{wang2019comparative, peukert2018direct, newcomb2017nonequilibrium, moore2017fate, zhang2025minimized, xiangyuan2020combustion, zhou2013multitarget} and turbulence-chemistry interaction(TCI)\citep{borghi2021ted,hadadpour2023extended}. Choosing a right turbulent combustion model is crucial when we apply the finite rate method and multi-step mechanisms\citep{zhu2025homogeneous,lin2018skeletal}, but a definitive solution for predicting sub-grid TCI remains elusive. TCI can be modeled explicitly or implicitly, the former includes models such as the partial premixing (PaSR) model\citep{ferrarotti2019role} while the latter involves direct modeling of finite rate chemistry and its extension\citep{thabari2024assessment}. Large eddy simulation (LES) combined with chemical reaction mechanism has been widely used in turbulent combustion research\citep{zettervall2017small,wehrfritz2016large}, but it still faces a series of challenges. The complex and nonlinear nature of turbulent flow and chemical reactions presents challenges for sub-grid scale (SGS) models in accurately handling molecular diffusion and mixing. Many studies aim to improve SGS models for better prediction accuracy, but their performance varies significantly depending on the specific turbulent combustion case\citep{rutland2011large}. Accurate calculation of chemical source terms (\emph{i.e.}, the sum of individual reaction rates of a species over all steps in the reaction mechanism) is vital in turbulent combustion simulations, and smaller grid sizes and direct numerical simulation (DNS) improve accuracy but increase computational costs.

The calculation of the reaction rate of each species is the basis of the numerical simulation of combustion. Advances in computer technology enable more complex multi-step mechanisms and higher species resolution, but computational costs still pose challenges in resolving chemical kinetics. Similar to the SGS turbulence models, researchers aim to incorporate turbulent characteristic parameters, as seen in the eddy-dissipation-concept (EDC) model\citep{thabari2024assessment,gran1996numerical} to consider the influence of small eddy on the reaction rate in the grid. Species transport equations and ordinary differential equations (ODE) for reaction mechanisms can only give mean concentrations in a grid, so people try to incorporate empirical turbulence parameters for a more accurate chemical source term. Turbulent combustion models address the interaction between turbulence and chemical kinetics at a small scale, utilizing models such as the EDC model, the flamelet model, the partially stirred reactor (PaSR) model\citep{ferrarotti2019role}, and the probability density function (PDF)\citep{borghi2021ted,hadadpour2023extended,pope1981monte}. These models construct the chemical source term \(\omega_{i}\) for solving the species transport equation,
\begin{equation}
\label{eq4}
    \frac{\partial}{\partial t}\left( \rho Y_{i} \right) + \nabla \cdot \left( \rho vY_{i} \right) = \nabla \cdot \left( \Gamma_{i}\nabla Y_{i} \right) + \omega_{i}
\end{equation}
where \(v\) is the velocity and \(\Gamma_{i}\) the diffusion coefficient. The key to solving Eq.(\ref{eq4}) is to obtain \(\omega_{i}\) by solving the ODE of reaction mechanisms. However, the treatment of both chemical kinetics and the species transport equation can only give the mean concentrations\(\{{\overline{c}}_{i}\}\) of the grids. If one simply applies \(\{{\overline{c}}_{i}\}\) to calculate \(\omega_{i}\), the reaction rate will generally be overestimated.

EDC uses an effective volume fraction to correct the chemical source term of the mean concentration. It is considered that the reaction only occurs in the fine structure, and the chemical source term is expressed as\citep{gran1996numerical}
\begin{equation}
    \omega_{i} = \gamma\ \rho\frac{Y_{i}^{*} - Y_{i}^{0}}{\tau^{*}}
\end{equation}
where \(Y_{i}^{*}\) and \(Y_{i}^{0}\) are the mass fraction of species \(i\) before and after the characteristic time scale \(\tau^{*}\). The factor \(\gamma\ \)takes a role in volume correction and is determined by parameters such as kinematic viscosity, turbulent kinetic energy, and turbulent kinetic energy dissipation rate\citep{shamooni2020priori}.

PaSR suggests that the fluid in the reaction zone is partially stirred rather than completely mixed, leading to areas with adequate stirring for reactant contact and areas with insufficient stirring\citep{wu2019flame} The reaction rate is influenced by both chemical reactions and turbulent mixing.
\begin{equation}
    \omega_{i} = \kappa\rho\frac{Y_{i}^{*} - Y_{i}^{0}}{{\ \tau}^{*}}
\end{equation}

Similar to the EDC model, PaSR introduces a factor of time correction which is expressed as the relationship between chemical time scale \(t_{c}\) and mixed one \(t_{mix}\), \emph{i.e.}
\begin{equation}
    \kappa = \frac{t_{c}}{t_{c} + t_{mix}}
\end{equation}

A difficulty in applying PaSR model is the determination of \(t_{c}\). In particular for a multi-step mechanism one cannot even define this quantity.

A more popular model for TCI is the so-called PDF\citep{borghi2021ted,pope1981monte} which establishes the transport equations of scalars such as species concentration and mixing fraction. PDF has undergone continuous improvement in recent years\citep{zhong2025flamelet, un2025stochastic}. This model typically employs random scalars (concentration) for statistical averaging in calculating \(\omega_{i}\). However, it requires introducing turbulence parameters for solving the transport equation, making PDF calculations generally expensive.

The scale of chemical reactions is much smaller than that of small turbulence vortices which contain a vast number of molecules. If we consider the rate constant \emph{k} in Eq.(\ref{eq1}) as fixed or a proper mean value, for example an isothermal system, the rate of formation or consumption of a species in a fixed volume \emph{V} depends on species concentrations, without explicit involving the turbulence parameters, while turbulence impacts the reaction rate by altering the temperature, pressure, and concentration distributions. Here, we first derive a generalized rate law (GRL) for a diffusion non-equilibrium system and subsequently present the expression for \(\omega_{i}\) given a specific concentration distribution of species. This approach allows the derivation of \(\omega_{i}\) without introducing turbulence parameters, thereby facilitating the eventual decoupling between turbulence and chemical kinetics in turbulent combustion simulations.

\section{Theory and method}
\label{sec:tam}
In practice, a combustion reaction mechanism is generally composed of a list of ``elementary'' steps, and a reaction is reversible in principle. In according with the law of mass action, the number of molecules in forward or backward reaction is generally not more than 2, while occasions of three particles colliding and reacting in space at the same time are rare\citep{atkins2023atkins}. Therefore, if we try to construct a mechanism with ``elementary'' steps, it usually includes only unimolecular or bimolecular reactions\citep{zhang2025minimized, xiangyuan2020combustion}. As shown in Eq.(\ref{eq1}), for a chemical reaction, the formation rate of species A (assuming the reactant) consists of two parts, the reaction rate constant \(k(T)\) and the product of concentrations \(c_{\mathrm A}^{\alpha}c_{\mathrm B}^{\beta}\cdots\). Except for a fully premixed and isothermal system, the temperature and species concentration have spatial distributions in the grid, while the species transport equation and energy equation can only give the mean values \(\overline{T}\) and \(\left\{ {\overline{c}}_{i} \right\}(i = 1\sim N)\) of the grids. In the following, we confine our discussions to the species concentrations, by assuming the invariant \emph{k} by assuming an isothermal system or an mean value \(\overline{k}\).

\subsection{A generalized rate law for inhomogeneous species distributions}

For a box with a volume \emph{V} in which all the species distribute homogeneously, the reaction rate can be expressed by the well-known rate law as shown by Eq.(\ref{eq1}). However, for an inhomogeneous system, Eq.(\ref{eq1}) using mean concentrations will cause errors. For the convenience of comparison, in the following we define the zeroth-order precision of reaction rate by using the mean concentrations for the mean reaction rate in Eq.(\ref{eq1}). It is obvious that the zero-order precision can give accurate value for homogenous systems.

Evaluating the reaction rate remains a challenge for turbulent combustion because the molecules do not reach the diffusion equilibrium unless perfectly mixed. For this purpose, one can take the measure of dividing the box (or the grid) with a volume \emph{V} into \emph{m} sub-grids. If the size is small enough, Eq.(\ref{eq1}) holds true in sub-grids. We assume \(V_{j}\) the volume sub-grid \emph{j}, and \(n_{A,j}\) the molar number of species A, we have the definition of the mean concentration as follows,
\begin{equation}
    {\overline{c}}_{\rm A,j} = \frac{n_{\rm A,j}}{V_{j}}
\end{equation}
The total molar number of A in the grid reads
\begin{equation}
    n_{\mathrm{A}} = \sum_{j = 1}^{m}{n_{\mathrm{A},j} = \ \sum_{j = 1}^{m}{{\overline{c}}_{\mathrm{A},j}V_{j}}}
\end{equation}
Taking the limit, we have
\begin{equation}
\label{eq10}
    n_{\mathrm{A}} = \lim_{V_{j} \rightarrow 0}\sum_{j = 1}^{m}{{\overline{c}}_{\mathrm{A},j}V_{j}} = \int_{}^{}{c_{\mathrm{A}}dv}
\end{equation}
with
\begin{equation}
\label{eq11}
    c_{\mathrm{A}}(x,y,z,t) = \lim_{V_{j} \rightarrow 0}\frac{n_{\mathrm{A},j}}{V_{j}}
\end{equation}
being the spatial concentration distribution. In this way, the mean reaction rate of grid can be written as
\begin{equation}
\label{eq12}
    \frac{d{\overline{c}}_{\mathrm{A}}}{dt} = \frac{dn_{\mathrm{A}}}{Vdt} = \frac{d}{Vdt}\int_{}^{}{c_{\mathrm{A}}dv} = \frac{1}{V}\int_{}^{}{\frac{dc_{\mathrm{A}}}{dt}dv}
\end{equation}

For the sake of brevity, the variables \(x,y,z\) and \(t\) in Eq.(\ref{eq11}) are omitted. When we take the limit of \(V_{j} \rightarrow 0\), Eq.(\ref{eq1}) becomes valid. Substituting Eq.(\ref{eq1}) into Eq.(\ref{eq12}) we have
\begin{equation}
\label{eq13}
    {\widehat{R}}_{\mathrm{A}} = \frac{d{\overline{c}}_{\mathrm{A}}}{dt} = - \frac{k}{V}\int_{}^{}{c_{\mathrm{A}}^{\alpha}c_{\mathrm{B}}^{\beta}\cdots dv}
\end{equation}
Here we assume an isothermal system or \(k\) a mean value of rate constant in value \emph{V}. Eq.(\ref{eq13}) is the new form of rate law for inhomogeneous system that we call GRL. For a homogeneous system, Eq.(\ref{eq13}) reduces to the standard form as shown by Eq.(\ref{eq1}) since concentrations become constant. Eq.(\ref{eq13}) indicates that if the proper spatial distributions of species concentrations within the grid are given, the reaction rates \(\frac{d{\overline{c}}_{A}}{dt}\) and further the chemical source terms can be obtained by performing the spatial integration over the product of concentrations, without the need of turbulent parameters explicitly involved. For a multi-step mechanism, \(\omega_{i}\) is given by summing the reaction rates of Eq.(\ref{eq13}) over all the formation and consumption steps for species, as that shown by Eq.(\ref{eq3}). Based on Eq.(\ref{eq13}) the calculation of mean reaction rates, further the chemical source terms in inhomogeneous reaction system such turbulent combustion can be worked out by the sub-grid concentration modeling, rather than sub-grid TCI.

\subsection{Reaction rates for typical reactions}
\subsubsection{First-order reaction}
In the following, we only discuss the one-way reaction, while the reverse reaction can be treated in the similar way. The first-order reaction is as follows,
\begin{equation}
    \mathrm{A}\ \ \ \overset{k}{\operatorname*{\operatorname*{\to}}} \ \ \ \mathrm{P}
\end{equation}
Cracking and isomerization of fuel molecules under high-pressure conditions belong to this kind of reactions, and the reaction rate for a homogeneous system is expressed as
\begin{equation}
    \frac{d{\overline{c}}_{\mathrm{A}}}{dt} = - k{\overline{c}}_{\mathrm{A}}
\end{equation}
When taking the inhomogeneous system into account, we consider the spatial distribution \(c_{\mathrm{A}} = c_{\mathrm{A}}(x,y,z,t)\), applying Eq.(\ref{eq13}) we have
\begin{equation}
    \frac{d{\overline{c}}_{\mathrm{A}}}{dt} = - \frac{k}{V}\int_{}^{}{c_{\mathrm{A}}dv} = - k{\bar{c}}_{\mathrm{A}}
\end{equation}
It can be seen that the first-order reaction rate is only related to the total number of species A in the grid and has nothing to do with the spatial distribution of \(c_{\mathrm{A}}\), and hence the reaction rate can be given exactly with the mean species concentration, without need of the introduction of turbulence parameters.
\subsubsection{Second-order reaction}
We consider a second-order reaction
\begin{equation}
    \mathrm{A} + \mathrm{B}\ \ \ \overset{k}{\rightarrow}\ \ \ \mathrm{P}
\end{equation}
Assuming the spatial distributions of species A and B as \(c_{\mathrm{A}}\) and \(c_{\mathrm{B}}\), the reaction rate can be written as
\begin{equation}
\label{eq18}
    \frac{d{\bar{c}}_{A}}{dt} = - \frac{k}{V}\int_{}^{}{c_{\mathrm{A}}c_{\mathrm{A}}dv}
\end{equation}
according to Eq.(\ref{eq13}). Such reactions in combustion include oxidation, free radical recombination, and so on. The spatial integration of \(c_{\mathrm{A}}c_{\mathrm{B}}\) in Eq.(\ref{eq18}) occupies the central position in our new theory for turbulence-chemistry decoupling.

If one species takes the homogeneous distribution, we have
\begin{equation}
\label{eq19}
    \frac{d{\bar{c}}_{\mathrm{A}}}{dt} = - \frac{k}{V}{\overline{c}}_{\mathrm{A}}\int_{}^{}{c_{\mathrm{B}}dv} = - \frac{k}{V}{\overline{c}}_{\mathrm{A}}{\overline{c}}_{\mathrm{B}}
\end{equation}
Eq.(\ref{eq19}) indicates that zero-order precision give the exact reaction rate. However, in the cases that both \(c_{\mathrm{A}}\) and \(c_{\mathrm{A}}\) are inhomogeneous, the reaction rate shown by Eq.(\ref{eq18}) needs the spatial integration of \(c_{\mathrm{A}}c_{\mathrm{B}}\) by introducing the spatial concentration distributions within the grid. Results for reaction rate for several types of concentration distributions are given in the next section.
\subsubsection{Lindemann unimolecular reaction}
In order to describe the pressure effect on rate constant \emph{k}, Lindemann proposed the following unimolecular reaction mechanism\citep{atkins2023atkins}
\begin{equation}
    \mathrm{A} + \mathrm{M} \; 
\underset{k_{-1}}{\overset{k_1}{\rightleftharpoons}} \;
\mathrm{A}^{*} + \mathrm{M}
\end{equation}
\begin{equation}
    \mathrm{A}^*\overset{k_2}{\operatorname*{\operatorname*{\to}}}\mathrm{P}
\end{equation}
In terms of the quasi-steady-state approximation, the formation rate of P is derived to the following form for a homogeneous system, \emph{i.e.}
\begin{equation}
    \frac{dc_{\mathrm{P}}}{dt} = \frac{k_{1}k_{2}c_{\mathrm{M}}c_{\mathrm{A}}}{k_{2} + k_{- 1}c_{\mathrm{M}}}
\end{equation}

Here \(c_{\mathrm{M}}\) is the sum of concentrations over all the species. We consider the simple one-dimensional situation and approximate \(c_{\mathrm{M}}\) constant, we have
\begin{equation}
\label{eq23}
    \frac{d{\bar{c}}_{\mathrm{P}}}{dt} = \frac{k_{1}k_{2}{\bar{c}}_{\mathrm{M}}}{k_{2} + k_{- 1}{\bar{c}}_{\mathrm{M}}}\int_{0}^{1}{c_{\mathrm{A}}dx} = \frac{k_{1}k_{2}{\bar{c}}_{\mathrm{M}}}{k_{2} + k_{- 1}{\bar{c}}_{\mathrm{M}}}{\bar{c}}_{\mathrm{A}}
\end{equation}
The Lindemann unimolecular mechanism is commonly used for cracking reactions in combustion. Eq.(\ref{eq23}) shows the first-order reaction feature.

\subsubsection{Trimolecular reaction and global mechanism}
Trimolecular reactions in reaction mechanisms of combustion generally contain a species M, and its concentration \(c_{\mathrm{M}}\) generally stands for the third-body concentration and is the sum of all the species. Similar to Lindemann mechanism, we can approximate \(c_{\mathrm{M}}\) constant and hence the rate reduces to the second-order case. The spatial integration for second-order reactions is valid for the reduced trimolecular reactions, \emph{i.e.}
\begin{equation}
    \frac{d{\bar{c}}_{\mathrm{A}}}{dt} = - \frac{k}{V}\int_{}^{}{c_{\mathrm{M}}c_{\mathrm{A}}c_{\mathrm{B}}dv} = - \frac{k{\bar{c}}_{\mathrm{M}}}{V}\int_{}^{}{c_{\mathrm{A}}c_{\mathrm{B}}dv}
\end{equation}

For the purpose practical combustion simulation, there are some global mechanisms which take the rate law form of Eq.(\ref{eq1}). Due to the complexity of chemical reactions, the reaction orders are often determined by experimental fittings and indexes \(\alpha\) and \(\beta\) are generally not integers. In such cases, the spatial integration of \(c_{\mathrm{A}}^{\alpha}c_{\mathrm{B}}^{\beta}\) will become complicated in particular in the non-integer index cases. Factually, owing to the empirical nature of global mechanisms, we are permitted to perform the calculation with lower accuracy, for example we can obtain the reaction rate by discretizing the power products, instead of the complicated spatial integrations.

\section{Results and discussions}
\label{sec:rad}
From the derivation above, the second-order reaction occupies the central position in calculating the rate of inhomogeneous systems. Here we suppose several types of concentration distributions in one-dimensional grids. In practical calculations, mean concentrations \({\{\overline{c}}_{i}\}\) can be obtained by solving the species transport equation (Eq.(\ref{eq4})). In order for the gain of concentration distribution in grid G(0), we take mean concentrations \({\overline{c}}_{\mathrm{A},0}\), \({\overline{c}}_{\mathrm{A},x -}\)and \({\overline{c}}_{\mathrm{A},x +}\) for grid G(0) and adjacent grids G(\emph{x}-) and G(\emph{x}+) into account. Here we choose the simplest linear variation of concentrations for our purpose. By the way one can of course choose more complicated concentration distributions according to for example the influences of turbulence.

\begin{figure}
	\centering
	\includegraphics[width=0.8\textwidth]{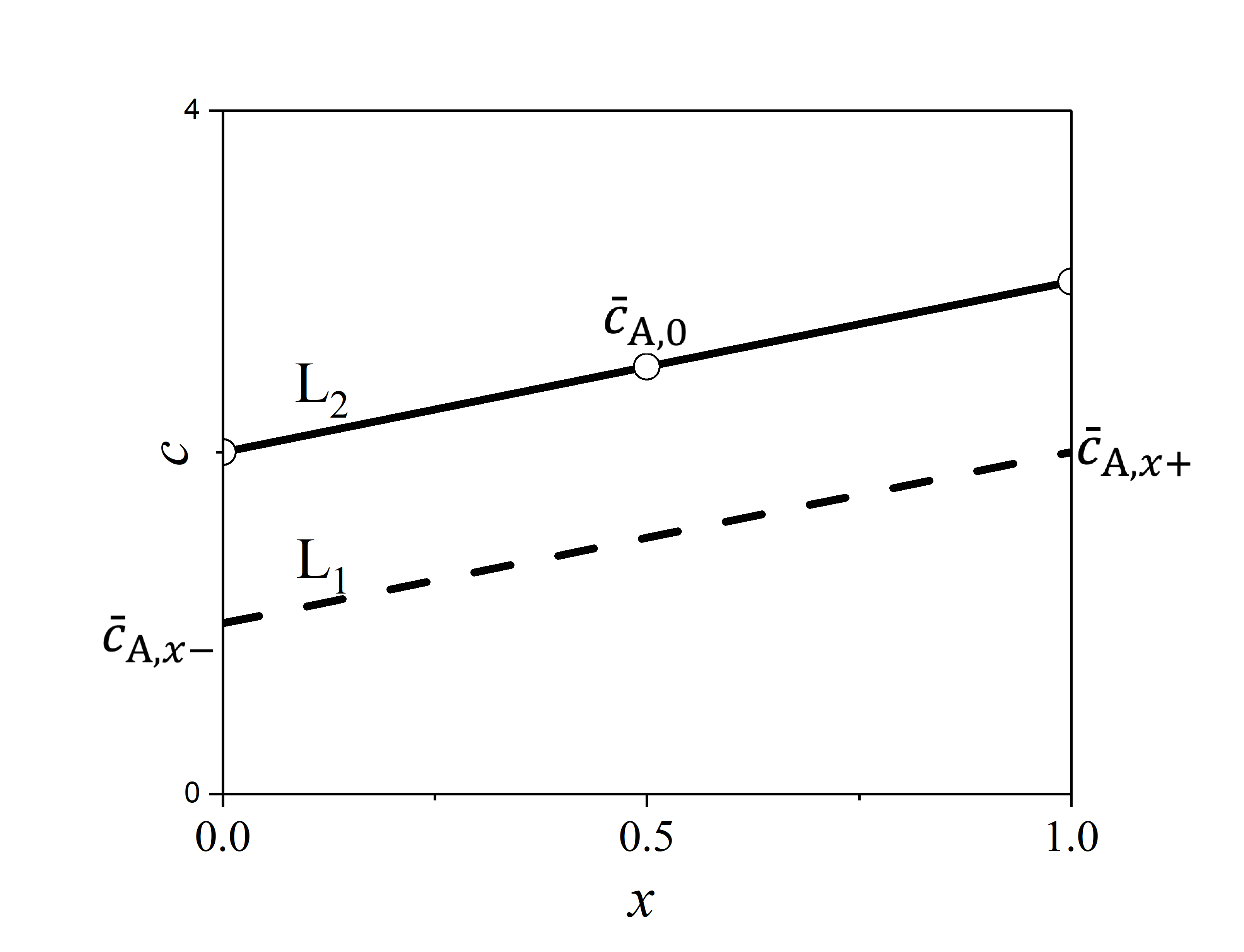}
	\caption{Diagram of one-dimensional linear concentration
distributions.}
	\label{fig:fig1}
\end{figure}

By assuming that the concentrations of species change linearly against \emph{x}-axis, we prefer to take \({\overline{c}}_{\mathrm{A},x-}\)and \({\overline{c}}_{\mathrm{A},x +}\) of the adjacent grids G(\emph{x}-) and G(\emph{x}+) as the boundary values of grid G(0), thus we obtain a linear distribution \(c_{\mathrm{A}}\) through the grid G(0) as the dashed line L\textsubscript{1} in Fig.\ref{fig:fig1}. It is noticed that the concentration distribution (dashed line) connects the boundaries of the grid G(0) and the intercepts are \({\overline{c}}_{\mathrm{A},x -}\) and \({\overline{c}}_{\mathrm{A},x +}\). To ensure the conservation of particle numbers, as shown by Eq.(\ref{eq10}), the dashed line L\textsubscript{1} is translated to the L\textsubscript{2} (the solid line in Fig.\ref{fig:fig1}) so that \(c_{\mathrm{A}} = {\overline{c}}_{\mathrm{A},0}\) at \emph{x} =0.5. Up to this stage, we use L\textsubscript{2} to model the spatial distribution of species A. Similarly, a linear distribution of species B can also be defined. Finally, we express the concentration distributions of a couple of species as follows,
\begin{equation}
\label{eq25}
    c_{\mathrm{A},x} = \left( {\overline{c}}_{\mathrm{A},x +} - {\overline{c}}_{\mathrm{A},x -} \right)x + {\overline{c}}_{\mathrm{A},0} - \frac{\left( {\overline{c}}_{\mathrm{A},x +} - {\overline{c}}_{\mathrm{A},x -} \right)}{2}
\end{equation}

\begin{equation}
\label{eq26}
    \ c_{\mathrm{B},x} = \left( {\overline{c}}_{\mathrm{B},x +} - {\overline{c}}_{\mathrm{B},x -} \right)x + {\overline{c}}_{\mathrm{B},0} - \frac{\left( {\overline{c}}_{\mathrm{B},x +} - {\overline{c}}_{\mathrm{B},x -} \right)}{2}
\end{equation}

Assuming a homogeneous constant \(k\) in the grid, we perform the spatial integration along \emph{x}-axis in grid G(0) by substituting Eqs.(\ref{eq25}) and (\ref{eq26}) into Eq.(\ref{eq18}), we have
\begin{equation}
\label{eq27}
    \frac{d{\bar{c}}_{\mathrm{A}}}{dt} = - k\int_{0}^{1}{c_{\mathrm{A},x}c_{\mathrm{B},x}dx} = - k\left\lbrack {\overline{c}}_{\mathrm{A},0}{\overline{c}}_{\mathrm{B},0} + \frac{{(\overline{c}}_{\mathrm{A},x +} - {\overline{c}}_{\mathrm{A},x -}){({\overline{c}}_{\mathrm{B},x +} - \overline{c}}_{\mathrm{B},x -})}{12} \right\rbrack
\end{equation}

Compared with the zero-order precision of reaction rate, we can define that the reaction rate with linear distribution assumption as shown in Eq.(\ref{eq27}) has the first-order precision, but such definitions are arguable and qualitative. Considering the existence of turbulence in the grid, a more realistic concentration distribution can be modeled to give more reliable results.

From Eq.(\ref{eq27}), one can also see that if
\({\overline{c}}_{\mathrm{A},x +} = {\overline{c}}_{\mathrm{A},x -}\) or
\({{\overline{c}}_{\mathrm{B},x +} = \overline{c}}_{\mathrm{B},x -}\),
\(\frac{d{\bar{c}}_{\mathrm{A}}}{dt}\) reduces to the case of homogeneous cases, \emph{i.e}.,
\begin{equation}
    \frac{d{\bar{c}}_{\mathrm{A}}}{dt} = - k{\overline{c}}_{\mathrm{A},0}{\overline{c}}_{\mathrm{B},0}
\end{equation}

Considering \emph{m} degrees of freedom in space, we can take more adjacent grids into account. Similar to the one-dimensional situation, the reaction rate of species A in grid G(0) can be approximated as
\begin{equation}
\label{eq29}
    \frac{d{\bar{c}}_{\mathrm{A}}}{dt} = - k\lbrack{\overline{c}}_{\mathrm{A},0}{\overline{c}}_{\mathrm{B},0} + \frac{1}{m}\sum_{i = 1}^{m}\frac{{(\overline{c}}_{\mathrm{A},i +} - {\overline{c}}_{\mathrm{A},i -}){({\overline{c}}_{\mathrm{B},i +} - \overline{c}}_{\mathrm{B},i -})}{12}\rbrack
\end{equation}

Eqs.(\ref{eq27}) and (\ref{eq29}) indicate that the reaction rate includes the zero-order team \(-k{\overline{c}}_{\mathrm{A},0}{\overline{c}}_{\mathrm{B},0}\) and the first-order term. In the above treatment, \({\overline{c}}_{\mathrm{A},i-}\) and \({\overline{c}}_{\mathrm{A},i +}\) of grids G(\emph{i}-) and G(\emph{i+}) have been taken as the boundary values of grid G(0). The slope of the distribution line in grid G(0) obtained by this method is \({\overline{c}}_{\mathrm{A},i +} - {\overline{c}}_{\mathrm{A},i-}\). One can make the different choice, for example the boundary values \(\frac{{\overline{c}}_{\mathrm{A},x-} + {\overline{c}}_{\mathrm{A},0}}{2}\) and \(\frac{{\overline{c}}_{\mathrm{A},0} + {\overline{c}}_{\mathrm{A},x+}}{2}\), to modeling the concentration distribution.

If the distribution has a more complicated form such as the case of small eddies within the grid, further consideration for concentration distributions should be taken. Two non-linier distributions of concentrations will be discussed in the following.

We take a second-order reaction as an example for the calculation of reaction rates. Considering the diffusion combustion of H\textsubscript{2} and O\textsubscript{2} under one-dimensional conditions we discuss the reaction rate of the elementary second-order reaction, namely
\begin{equation}
    \rm{H_{2} + O_{2}} \rightarrow \rm{HO}_{2} + \rm H
\end{equation}

\begin{figure}
	\centering
	\includegraphics[width=0.8\textwidth]{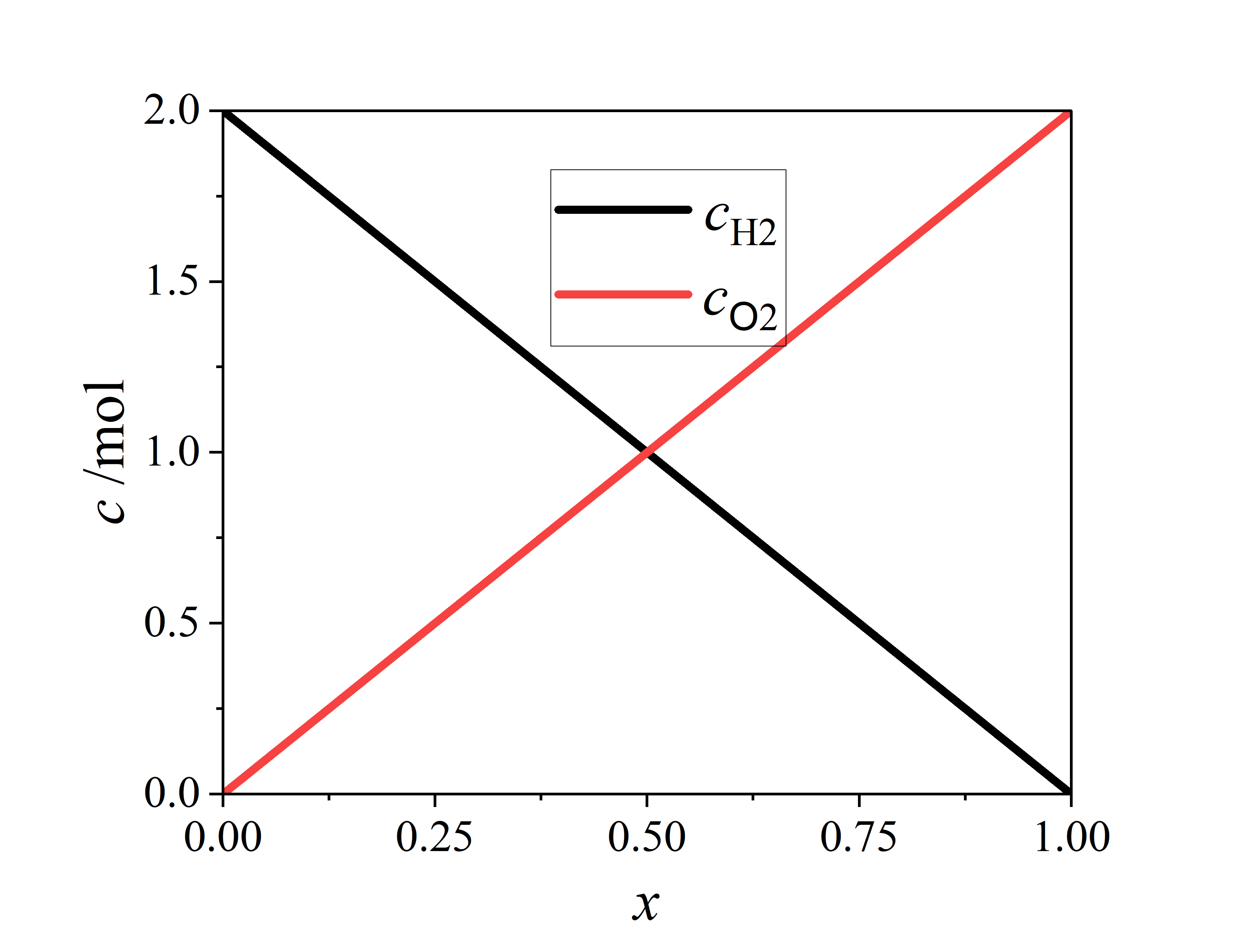}
	\caption{Linear concentration distributions of H\textsubscript{2}
and O\textsubscript{2} in grid.}
	\label{fig:fig2}
\end{figure}

We choose the straight line, the sine function, and the piecewise distributions in one-dimensional grid. As shown in Fig.\ref{fig:fig2}, the oxidizer and fuel nozzles are assumed to separate by a distance 1, and both H\textsubscript{2} and O\textsubscript{2} are 1 mol in total in this interval.

For the case of straight line, we assume that the concentration of H\textsubscript{2} decreases linearly from 2~mol at \emph{x}=0 to 0 at \emph{x =}1, and on the contrary O\textsubscript{2} from 0 to 2mol (Fig.\ref{fig:fig2}). The mean concentrations of both H\textsubscript{2} and O\textsubscript{2} are \(1mol\). As shown in Fig.\ref{fig:fig2}, the linear concentrations of H\textsubscript{2} and O\textsubscript{2} reads
\begin{equation}
\label{eq31}
    c_{\rm{H_2}} = - 2x + 2, c_{\rm{O_2}} = 2x
\end{equation}

Applying the law of mass action to Eq.(\ref{eq31}), the generation rate of H\textsubscript{2} of zero-order precision in the one-grid case is given by
\begin{equation}
\label{eq32}
    \frac{d{\bar{c}}_{\rm{H}_2}}{dt}(1) = - k{\overline{c}}_{\rm{H}_2}{\overline{c}}_{\rm{O}_2} = - k{\ mol}^{2}
\end{equation}

If the grid is further divided into 10 sub-grids, with the grid length being 1/10, the mean concentrations of H\textsubscript{2} and O\textsubscript{2} in the grids have been given in Table \ref{tab:table}. The sum of the products of the mean concentrations reads
\begin{equation}
    \frac{d{\bar{c}}_{\rm{H}_2}}{dt}(10) = - \frac{k}{10}\sum_{j = 1}^{10}{{\overline{c}}_{\rm{H}_2,j}{\overline{c}}_{\rm{O}_2,j}} = - 0.67~k\ {mol}^{2}
\end{equation}

\begin{table}
	\caption{Mean concentrations of H\textsubscript{2} and O\textsubscript{2}
in 10 sub-grids}
	\centering
	\begin{tabular}{ccccccccccc}
		\toprule
		Grid no. & 1& 2& 3& 4& 5& 6& 7& 8& 9& 10  \\
		\midrule
		\({\overline{c}}_{\rm{H}_2}\) & 1.9 & 1.7 & 1.5 & 1.3 & 1.1 & 0.9 & 0.7 & 0.5 & 0.3 & 0.1 \\
		\({\overline{c}}_{\rm{O}_2}\) & 0.1 & 0.3 & 0.5 & 0.7 & 0.9 & 1.1 & 1.3 & 1.5& 1.7 & 1.9 \\
		\bottomrule
	\end{tabular}
	\label{tab:table}
\end{table}

Furthermore, when divided into 100 sub-grids, we have \(\frac{d{\bar{c}}_{\rm{H}_2}}{dt}(100) = - 0.6667k\ {mol}^{2}\). Finally, supposing the linear distributions as shown by Eq.(\ref{eq31}), the spatial integration by Eq.(\ref{eq18}) gives
\begin{equation}
\label{eq34}
    \frac{d{\bar{c}}_{\rm{H}_2}}{dt}(\infty) = - k\int_{0}^{1}{c_{\rm{H}_2}c_{\rm{O}_2}dx} = - 2k/3\ {mol}^{2}
\end{equation}

Eq.(\ref{eq34}) is the accurate reaction rate if the concentration distributions possess the exact strait line form. It is easy to see that as the grid is subdivided, the zero-order precision reaction rate quickly approaches the accurate reaction rate \(- 2k/3\ {mol}^{2}\). If we take \(\frac{d{\bar{c}}_{\rm{H}_2}}{dt}(\infty)\) as the accurate value, the error caused by the approximation of zero-order precision is given by
\begin{equation}
    \delta = \lbrack\frac{d{\bar{c}}_{\rm{H}_2}}{dt}(1) - \frac{d{\bar{c}}_{\rm{H}_2}}{dt}(\infty)\rbrack/\frac{d{\bar{c}}_{\rm{H}_2}}{dt}(\infty) = \ 50\%
\end{equation}

\begin{figure}
	\centering
	\includegraphics[width=0.8\textwidth]{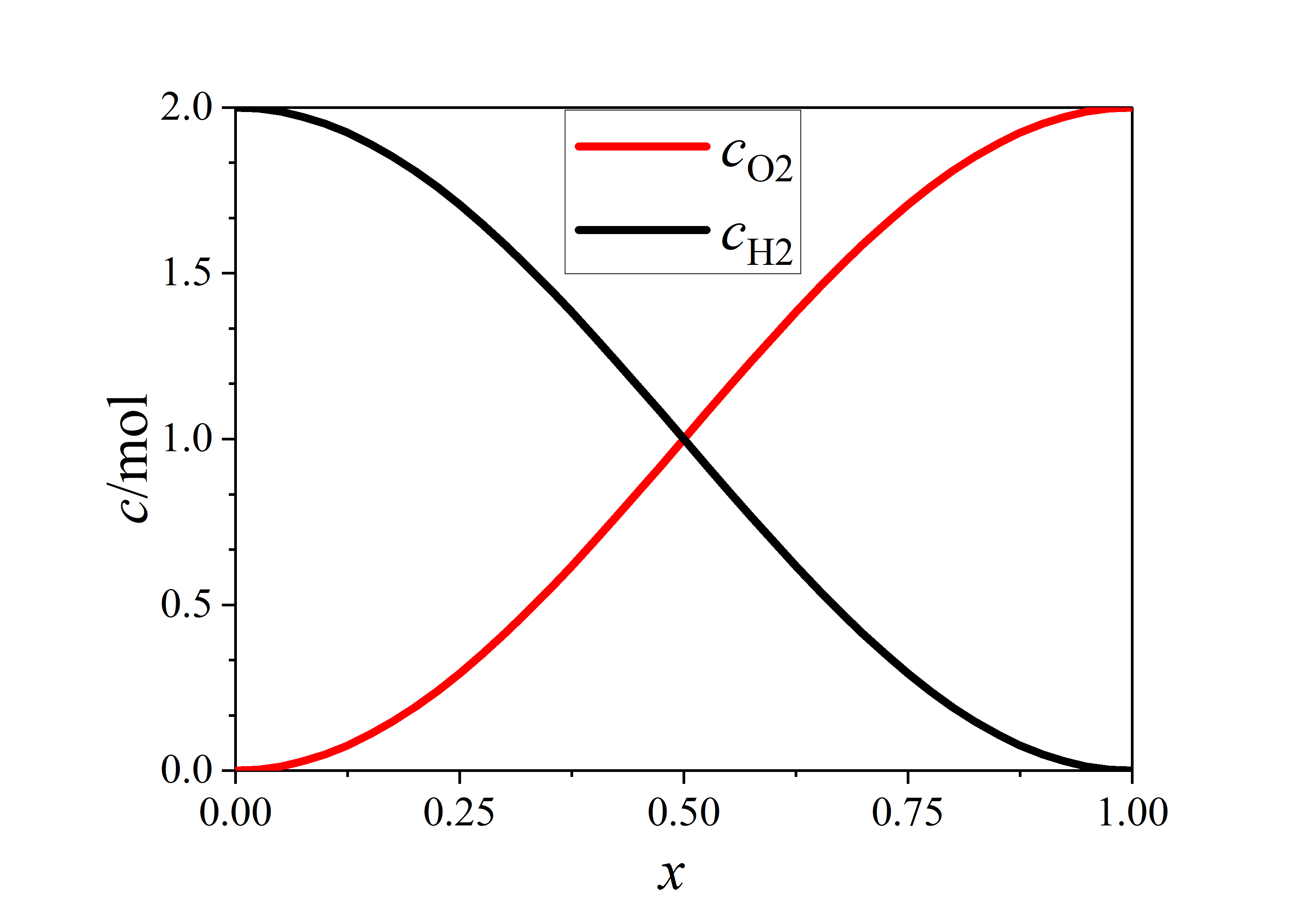}
	\caption{Sine function distributions of \(c_{\rm{O}_2}\) and \(c_{\rm{H}_2}\).}
	\label{fig:fig3}
\end{figure}

For the second case we suppose a distribution of sine function for \(c_{\rm{O}_2}\) and \(c_{\rm{H}_2}\) (Fig.\ref{fig:fig3}) in the one-dimensional grid with a length of\(\ 1\), \emph{i.e.}
\begin{equation}
\label{eq36}
    c_{\rm{H}_2} = 1 + \cos(\pi x),c_{\rm{O}_2} = 1 + \sin ( \pi x - \pi\text{/}2)
\end{equation}

The mean value \({\overline{c}}_{\rm{O}_2}\) can be determined as
\begin{equation}
    {\overline{c}}_{\rm{O}_2} = \int_{0}^{1}{\left\lbrack 1 + \sin\left( \pi x - \pi\text{/}2 \right) \right\rbrack\ dx = 1}mol
\end{equation}

and similarly, \({\overline{c}}_{\rm{H}_2} = 1\ mol\). The reaction rate by mean concentrations is as the same of Eq.(\ref{eq32}). Applying the spatial integration and substituting Eq.(\ref{eq36}) into Eq.(\ref{eq18}), we have
\begin{equation}
\label{eq38}
    \frac{d{\bar{c}}_{\rm{H}_2}}{dt}(\infty) = - k\int_{0}^{a}{\left\lbrack 1 + \sin\left( \pi x - \pi\text{/}2 \right) \right\rbrack\lbrack 1 + \cos(\pi x)\rbrack dx} = - k/2~{mol}^{2}
\end{equation}

Comprising with ``accurate'' spatial integration as given by Eq.(\ref{eq38}), the result of zero-order precision by Eq.(\ref{eq32}) overestimates the reaction rate by an error of \(100\%\). Comparing with the linear distribution, the sine function distribution reflects worse mixing performance and hence smaller reaction rate.

\begin{figure}
	\centering
	\includegraphics[width=0.8\textwidth]{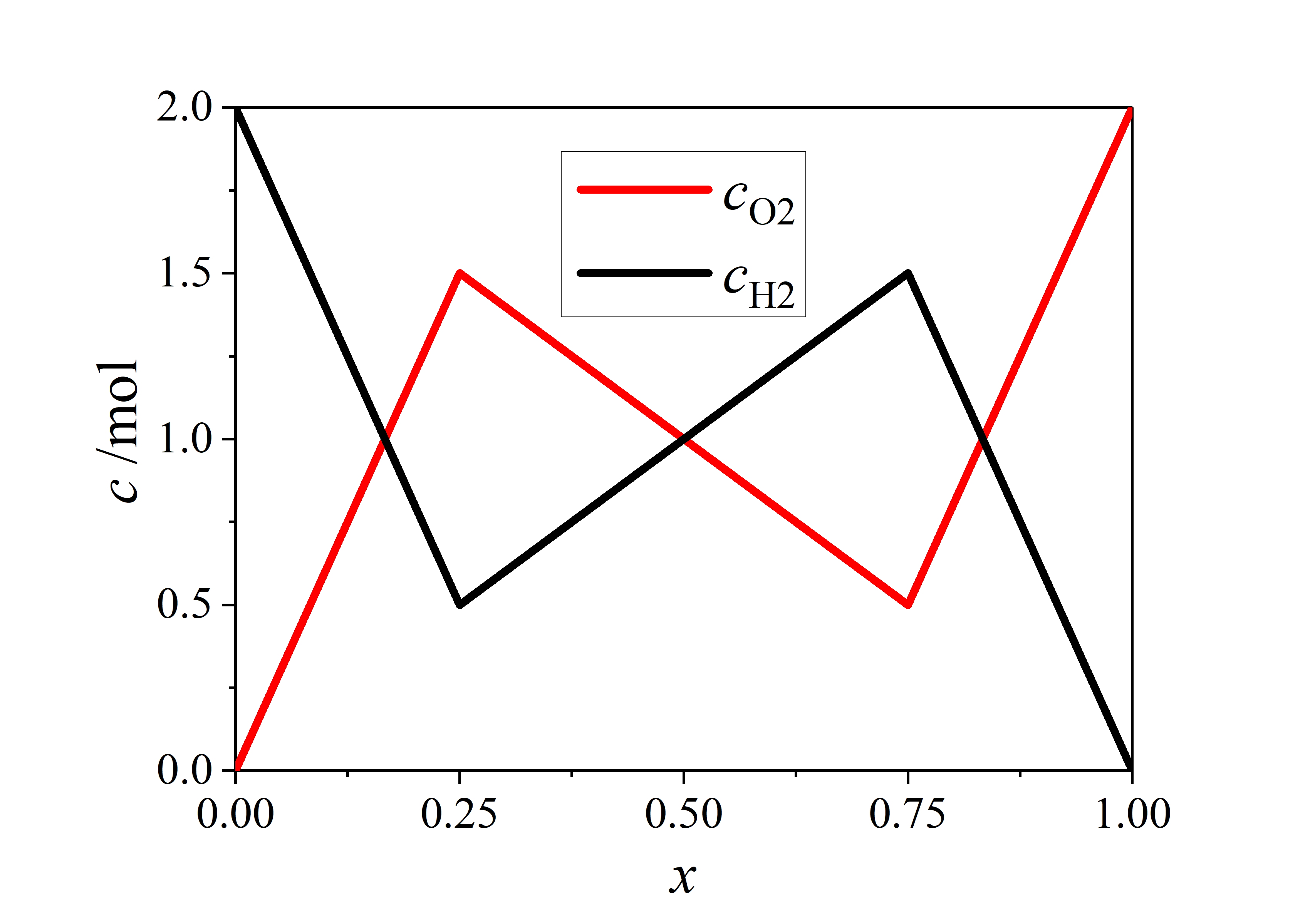}
	\caption{piecewise distributions of \(c_{\rm{O}_2}\) and \(c_{\rm{H}_2}\).}
	\label{fig:fig4}
\end{figure}

Further we suppose a piecewise distribution as shown in Fig.\ref{fig:fig4} with the values of concentrations being \({\overline{c}}_{\rm{H}_2} = {\overline{c}}_{\rm{O}_2} = 1~mol\), and the piecewise function of \(c_{\rm{O}_2}\) and \(c_{\rm{H}_2}\) reads
\begin{equation}
\begin{aligned}
c_{\mathrm{H}_2}(x) &= 2 - 6x, & c_{\mathrm{O}_2}(x) &= 6x, & x &\in [0, 1/4], \\
c_{\mathrm{H}_2}(x) &= 2x, & c_{\mathrm{O}_2}(x) &= 2 - 2x, & x &\in [1/4, 3/4], \\
c_{\mathrm{H}_2}(x) &= 6 - 6x, & c_{\mathrm{O}_2}(x) &= -4 + 6x, & x &\in [3/4, 1].
\end{aligned}
\end{equation}

Performing the spatial integration by Eq.(\ref{eq18}) we have \(\frac{d{\bar{c}}_{\rm{H}_2}}{dt}(\infty) = - 5k/6~{mol}^{2}\), while the mean concentration gives a value as the same as given by Eq.(\ref{eq32}). It is seen that in this case the zero-order precision produces an overestimation of 20\% for the reaction rate, being smaller than the linear distribution case as given by Eq.(\ref{eq34}). This means that the existence of eddies in the grid possibly causes the wrinkle distribution of species and makes the mixing improved, and thus the simple mean concentration can yield better reaction rates and chemical source terms. It should be noticed that the chemical source term does not explicitly include the turbulent parameters but how to correlate the species concentration distributions and the turbulence intensity remains an important aspect in future work.

\section{Conclusion}

In this paper, a generalized reaction rate law for inhomogeneous
concentration distributions of chemical species is derived by means of
spatial integration of the product of concentrations. This work provides
a way to calculate the reaction rates in the cases of diffusion
non-equilibrium systems. The method under discussion focuses on the
calculation of the reaction rate of chemical species in a multi-step
reaction mechanism. Considering several typical reaction types, the
first-, second-, and the third-order reactions, and the global
mechanisms, which are intrinsic to the reaction mechanisms, have been
taken as examples. The proposed methodology aims to circumvent the
inaccuracies that arise from the homogeneous mixing hypothesis. The
solution to the problem is to consider the spatial distribution of
species and to perform the spatial integration for the rate law. A
significant conclusion is that for the unimolecular reactions, and the
second-order reaction with one reactant perfectly mixed, the mean
concentrations give exact reaction rates, regardless of whether the
reactants are well mixed or not.

This study evaluates species concentration distribution models (straight
line, piecewise, and sine function) for a second-order reaction and
calculates reaction rates based on spatial integration. Results show
that the reaction rate by mean concentrations will produce the errors of
100\%, 50\%, and 20\%, comprising the accurate integrations based on the
concentration distribution assumptions of sine function, straight line,
and piecewise distribution, respectively. Results indicate that the
chemical species mix better, the mean concentrations give more reliable
reaction rates.

Inhomogeneous chemical species distributions lead to the difficulty of
calculating the chemical source term of species transport equations.
Since the vectors and scalars in the grid cannot be resolved, the
current turbulent and combustion models use the sub-grid approximation
to establish the species transport equations in turbulent combustion
simulations. The present work creates a new approach to the evaluation
of the chemical source term. According to the new formulism, the
reaction rates and the chemical source terms in combustion simulations
depend only on a proper spatial distribution of chemical species in the
grid or container, without explicit inclusion of turbulence parameters.

The present work only provides theoretical ideas, and readers are
encouraged to try their extension for example, more reliable
concentration distributions and other reaction types in applying the
newly established generalized rate law to turbulent reacting flows.

\section*{Acknowledgments}
This work is supported by National Natural Science Foundation of China
(No. T2441001).

\bibliographystyle{unsrtnat}
\bibliography{references}  

\begin{thebibliography}{27}
\providecommand{\natexlab}[1]{#1}
\providecommand{\url}[1]{\texttt{#1}}
\expandafter\ifx\csname urlstyle\endcsname\relax
  \providecommand{\doi}[1]{doi: #1}\else
  \providecommand{\doi}{doi: \begingroup \urlstyle{rm}\Url}\fi

\bibitem[Atkins et~al.(2023)Atkins, De~Paula, and Keeler]{atkins2023atkins}
Peter~William Atkins, Julio De~Paula, and James Keeler.
\newblock \emph{Atkins' physical chemistry}.
\newblock Oxford university press, 2023.

\bibitem[Jesudason(2009)]{jesudason2009internal}
Christopher~G Jesudason.
\newblock Internal thermodynamical equilibrium and consequences for rate law expressions for elementary chemical and physical reactions.
\newblock In \emph{AIP Conference Proceedings}, volume 1150, pages 132--139. American Institute of Physics, 2009.

\bibitem[Plutschack et~al.(2017)Plutschack, Pieber, Gilmore, and Seeberger]{plutschack2017hitchhiker}
Matthew~B Plutschack, Bartholom\"{a}us Pieber, Kerry Gilmore, and Peter~H Seeberger.
\newblock The hitchhiker's guide to flow chemistry.
\newblock \emph{Chemical reviews}, 117\penalty0 (18):\penalty0 11796--11893, 2017.

\bibitem[Egolfopoulos et~al.(2014)Egolfopoulos, Hansen, Ju, Kohse-H{\"o}inghaus, Law, and Qi]{egolfopoulos2014advances}
Fokion~N Egolfopoulos, Nils Hansen, Yiguang Ju, Katharina Kohse-H{\"o}inghaus, Chung~King Law, and Fei Qi.
\newblock Advances and challenges in laminar flame experiments and implications for combustion chemistry.
\newblock \emph{Progress in Energy and Combustion Science}, 43:\penalty0 36--67, 2014.

\bibitem[Westbrook et~al.(1977)Westbrook, Creighton, Lund, and Dryer]{westbrook1977numerical}
CK~Westbrook, J~Creighton, C~Lund, and FL~Dryer.
\newblock A numerical model of chemical kinetics of combustion in a turbulent flow reactor.
\newblock \emph{The Journal of Physical Chemistry}, 81\penalty0 (25):\penalty0 2542--2554, 1977.

\bibitem[Wang et~al.(2019)Wang, Sun, and Curran]{wang2019comparative}
Quan-De Wang, Yanjin Sun, and Henry~J Curran.
\newblock Comparative chemical kinetic analysis and skeletal mechanism generation for syngas combustion with no x chemistry.
\newblock \emph{Energy \& Fuels}, 34\penalty0 (1):\penalty0 949--964, 2019.

\bibitem[Peukert et~al.(2018)Peukert, Sela, Nativel, Herzler, Fikri, and Schulz]{peukert2018direct}
Sebastian Peukert, Paul Sela, Damien Nativel, J\"{u}rgen Herzler, Mustapha Fikri, and Christof Schulz.
\newblock Direct measurement of high-temperature rate constants of the thermal decomposition of dimethoxymethane, a shock tube and modeling study.
\newblock \emph{The Journal of Physical Chemistry A}, 122\penalty0 (38):\penalty0 7559--7571, 2018.

\bibitem[Newcomb et~al.(2017)Newcomb, Alaghemandi, and Green]{newcomb2017nonequilibrium}
Lucas~B Newcomb, Mohammad Alaghemandi, and Jason~R Green.
\newblock Nonequilibrium phase coexistence and criticality near the second explosion limit of hydrogen combustion.
\newblock \emph{The Journal of Chemical Physics}, 147\penalty0 (3), 2017.

\bibitem[Moore et~al.(2017)Moore, Turney, and Schaefer]{moore2017fate}
Kevin~B Moore, Justin~M Turney, and Henry~F Schaefer.
\newblock The fate of the tert-butyl radical in low-temperature autoignition reactions.
\newblock \emph{The Journal of Chemical Physics}, 146\penalty0 (19), 2017.

\bibitem[Zhang et~al.(2025)Zhang, Shi, Xia, Du, and Yao]{zhang2025minimized}
Wenhan Zhang, Lei Shi, Wenwen Xia, Yufan Du, and Li~Yao.
\newblock Minimized reaction network method for the construction of combustion reaction mechanism: Nh3/dme mixed combustion.
\newblock \emph{The Journal of Chemical Physics}, 163\penalty0 (7), 2025.

\bibitem[Xiangyuan et~al.(2020)Xiangyuan, Jiangtao, Yiwei, Juanqin, and Jingbo]{xiangyuan2020combustion}
Li~Xiangyuan, Shentu Jiangtao, Li~Yiwei, Li~Juanqin, and Wang Jingbo.
\newblock Combustion mechanism construction based on minimized reaction network: Hydrogen-oxygen combustion.
\newblock \emph{Gaodeng Xuexiao Huaxue Xuebao/Chemical Journal of Chinese Universities}, 41:\penalty0 772--779, 2020.

\bibitem[Zhou et~al.(2013)Zhou, Davis, and Skodje]{zhou2013multitarget}
Dingyu~DY Zhou, Michael~J Davis, and Rex~T Skodje.
\newblock Multitarget global sensitivity analysis of n-butanol combustion.
\newblock \emph{The Journal of Physical Chemistry A}, 117\penalty0 (17):\penalty0 3569--3584, 2013.

\bibitem[Borghi(2021)]{borghi2021ted}
R~Borghi.
\newblock Ted o'brien and turbulent combustion, 2021.

\bibitem[Hadadpour et~al.(2023)Hadadpour, Xu, Zhang, Bai, and Jangi]{hadadpour2023extended}
Ahmad Hadadpour, Shijie Xu, Yan Zhang, Xue-Song Bai, and Mehdi Jangi.
\newblock An extended fgm model with transported pdf for les of spray combustion.
\newblock \emph{Proceedings of the Combustion Institute}, 39\penalty0 (4):\penalty0 4889--4898, 2023.

\bibitem[Zhu et~al.(2025)Zhu, Li, Gao, Shi, Hu, and Liu]{zhu2025homogeneous}
Shichao Zhu, Pengfei Li, Yan Gao, Guodong Shi, Fan Hu, and Zhaohui Liu.
\newblock Homogeneous fuel-no mitigation during flameless oxy-combustion of ch4/nh3 mixtures.
\newblock \emph{Energy \& Fuels}, 39\penalty0 (6):\penalty0 3266--3279, 2025.

\bibitem[Lin and Lee(2018)]{lin2018skeletal}
Kuang~C Lin and Tzu-Wei Lee.
\newblock Skeletal mechanism of ethyl propionate oxidation for cfd modeling to predict experimental profiles of unsaturated products in a nonpremixed flame.
\newblock \emph{Energy \& Fuels}, 32\penalty0 (1):\penalty0 855--866, 2018.

\bibitem[Ferrarotti et~al.(2019)Ferrarotti, Li, and Parente]{ferrarotti2019role}
Marco Ferrarotti, Zhiyi Li, and Alessandro Parente.
\newblock On the role of mixing models in the simulation of mild combustion using finite-rate chemistry combustion models.
\newblock \emph{Proceedings of the combustion institute}, 37\penalty0 (4):\penalty0 4531--4538, 2019.

\bibitem[Thabari et~al.(2024)Thabari, Kruljevic, Maragkos, Snegirev, and Merci]{thabari2024assessment}
Jeri~At Thabari, Boris Kruljevic, Georgios Maragkos, Alexander Snegirev, and Bart Merci.
\newblock Assessment of the edc/finite rate chemistry approach towards predicting extinction in a turbulent buoyant diffusion flame.
\newblock \emph{Proceedings of the Combustion Institute}, 40\penalty0 (1-4):\penalty0 105602, 2024.

\bibitem[Zettervall et~al.(2017)Zettervall, Fureby, and Nilsson]{zettervall2017small}
Niklas Zettervall, Christer Fureby, and Elna~JK Nilsson.
\newblock Small skeletal kinetic reaction mechanism for ethylene--air combustion.
\newblock \emph{Energy \& Fuels}, 31\penalty0 (12):\penalty0 14138--14149, 2017.

\bibitem[Wehrfritz et~al.(2016)Wehrfritz, Kaario, Vuorinen, and Somers]{wehrfritz2016large}
Armin Wehrfritz, Ossi Kaario, Ville Vuorinen, and Bart Somers.
\newblock Large eddy simulation of n-dodecane spray flames using flamelet generated manifolds.
\newblock \emph{Combustion and Flame}, 167:\penalty0 113--131, 2016.

\bibitem[Rutland(2011)]{rutland2011large}
CJ~Rutland.
\newblock Large-eddy simulations for internal combustion engines--a review.
\newblock \emph{International Journal of Engine Research}, 12\penalty0 (5):\penalty0 421--451, 2011.

\bibitem[Gran and Magnussen(1996)]{gran1996numerical}
Inge~R Gran and Bj{\o}rn~F Magnussen.
\newblock A numerical study of a bluff-body stabilized diffusion flame. part 2. influence of combustion modeling and finite-rate chemistry.
\newblock \emph{Combustion Science and Technology}, 119\penalty0 (1-6):\penalty0 191--217, 1996.

\bibitem[Pope(1981)]{pope1981monte}
Stephen~B Pope.
\newblock A monte carlo method for the pdf equations of turbulent reactive flow.
\newblock \emph{Combustion Science and Technology}, 25\penalty0 (5\&6):\penalty0 159--174, 1981.

\bibitem[Shamooni et~al.(2020)Shamooni, Cuoci, Faravelli, and Sadiki]{shamooni2020priori}
Ali Shamooni, A~Cuoci, T~Faravelli, and A~Sadiki.
\newblock An a priori dns analysis of scale similarity based combustion models for les of non-premixed jet flames.
\newblock \emph{Flow, Turbulence and Combustion}, 104\penalty0 (2):\penalty0 605--624, 2020.

\bibitem[Wu et~al.(2019)Wu, Piao, Xie, and Ren]{wu2019flame}
Wantong Wu, Ying Piao, Qing Xie, and Zhuyin Ren.
\newblock Flame diagnostics with a conservative representation of chemical explosive mode analysis.
\newblock \emph{AIAA Journal}, 57\penalty0 (4):\penalty0 1355--1363, 2019.

\bibitem[Zhong et~al.(2025)Zhong, Xu, Weng, Cai, and Chen]{zhong2025flamelet}
Shenghui Zhong, Shijie Xu, Wubin Weng, Weiwei Cai, and Longfei Chen.
\newblock A flamelet-based eulerian transported pdf method for the modeling and simulation of supersonic combustion.
\newblock \emph{Combustion and Flame}, 272:\penalty0 113864, 2025.

\bibitem[Un and Navarro-Martinez(2025)]{un2025stochastic}
Tin-Hang Un and Salvador Navarro-Martinez.
\newblock Stochastic fields with adaptive mesh refinement for high-speed turbulent combustion.
\newblock \emph{Combustion and Flame}, 272:\penalty0 113897, 2025.

\end{thebibliography}






\end{document}